\documentclass[%
    reprint,
    prx,
    superscriptaddress,
    nofootinbib,
]{revtex4-2}
\usepackage{graphicx}
\usepackage{microtype}              
\usepackage[utf8]{inputenc}         
\usepackage[T1]{fontenc}            
\usepackage{newtxtext}              
\usepackage[smallerops]{newtxmath}  
\usepackage{bm}
\usepackage[dvipsnames]{xcolor}
\usepackage{hyperref}
\hypersetup{
    colorlinks=true,
    linkcolor=NavyBlue,
    citecolor=NavyBlue,
    urlcolor=NavyBlue
}

\renewcommand{\H}{E}
\newcommand{\iid}{\textit{iid}}
\DeclareMathOperator{\sech}{sech}
\DeclareMathOperator{\var}{var}
\DeclareMathOperator{\std}{std}
\newcommand{\para}[1]{\textbf{#1}}  
\newcommand{\CUNY}{\affiliation{%
    Initiative for the Theoretical Sciences,
    The Graduate Center, CUNY, 
    New York, New York 10016, USA%
    }%
}
\newcommand{\equalcontrib}{\thanks{%
    These authors contribute equally to this work.%
    }%
}
\begin{document}\frenchspacing
\title{Extrinsic vs Intrinsic Criticality in Systems with Many Components}
\author{Vudtiwat~Ngampruetikorn}
    \email{wave.ngamp@gmail.com}
    \CUNY
\author{Ilya~Nemenman}\equalcontrib
    \affiliation{%
    Department of Physics, Department of Biology, and 
    Initiative in Theory and Modeling of Living Systems,
    Emory University, Atlanta, Georgia 30322, USA%
    }%
\author{David~J.~Schwab}\equalcontrib
    \CUNY
\begin{abstract}
Biological systems with many components often exhibit seemingly critical behaviors, characterized by atypically large correlated fluctuations. Yet the underlying causes remain unclear. Here we define and examine two types of criticality. \emph{Intrinsic} criticality arises from interactions within the system which are fine-tuned to a critical point. \emph{Extrinsic} criticality, in contrast, emerges without fine tuning when observable degrees of freedom are coupled to unobserved fluctuating variables. We unify both types of criticality using the language of learning and information theory. We show that critical correlations, intrinsic or extrinsic, lead to diverging mutual information between two halves of the system, and are a feature of learning problems, in which the unobserved fluctuations are inferred from the observable degrees of freedom. We argue that extrinsic criticality is equivalent to standard inference, whereas intrinsic criticality describes \emph{fractional learning}, in which the amount to be learned depends on the system size. We show further that both types of criticality are on the same continuum, connected by a smooth crossover. In addition, we investigate the observability of Zipf's law, a power-law rank-frequency distribution often used as an empirical signature of criticality. We find that Zipf's law is a robust feature of extrinsic criticality but can be nontrivial to observe for some intrinsically critical systems, including critical mean-field models. We further demonstrate that models with global dynamics, such as oscillatory models, can produce observable Zipf's law without relying on either external fluctuations or fine tuning. Our findings suggest that while possible in theory, fine tuning is not the only, nor the most likely, explanation for the apparent ubiquity of criticality in biological systems with many components. Our work offers an alternative interpretation in which criticality, specifically extrinsic criticality, results from the adaptation of collective behavior to external stimuli. 
\end{abstract}
\maketitle

Life emerges from an intricate interplay among a large number of components, yet how it achieves such exquisite organization remains unexplained. Several aspects of this question fall within the domain of statistical physics, which studies the emergence of collective behaviors from the interaction between microscopic degrees of freedom. Perhaps the greatest success of statistical physics is in describing spontaneous transitions between two phases of matter such as liquid and gas. For a class of phase transitions, such as between ferromagnetic and paramagnetic states, the critical point, which separates the two phases, displays unique properties, absent from either phase. Many of these properties are relevant to biological function: scale invariance allows scaling up without the need for redesign~\cite{west:97,stevens:01,lee:07}, insensitivity to microscopic details can form a basis for robust behaviors~\cite{krotov:14,zhang:22}, and strong correlations between components appear useful for effective information propagation~\cite{beggs:08,chialvo:10,toyoizumi:11,mathijssen:19}.

A tantalizing question arises whether biology operates near a critical point \cite{mora:11,beggs:22}. This idea has a long history, see, e.g., Ref.~\cite{munoz:18}. However, it is not until recently that high-precision, simultaneous measurements of hundreds to thousands of components in biological systems allow quantitative empirical tests of the criticality hypothesis.

Modern quantitative biology experiments have indeed observed seemingly critical behaviors in many systems across scales, from amino acid sequences~\cite{mora:10} to spatiotemporal dynamics of gene expressions~\cite{nykter:08,krotov:14} to firing patterns of neurons~\cite{beggs2003neuronal,haimovici:13,tkacik:15,mora:15,kastner:15,meshulam:19} to velocity fluctuations in bird flocks~\cite{cavagna:10,bialek:12,bialek:14}. In these systems, correlations among the components and susceptibilities to perturbations often appear to diverge with the system size. Yet, this ubiquity is somewhat surprising, not least because equilibrium statistical physics tells us that criticality requires a hard to achieve fine tuning of models to a special point in their parameter space.

While biology may well be capable of fine tuning~\cite{attanasi:14,ma:19}, alternative explanations for the observed criticality exists~\cite{schwab:14,aitchison:17,touboul:17}. Some signatures of criticality arise without fine tuning when observable degrees of freedom are coupled to an unobserved fluctuating variable or variables~\cite{schwab:14,aitchison:17,morrell:20,morrell:23}. Provided that the number of fluctuating variables is relatively small and their fluctuations are sufficiently large, this latent fluctuation needs not depend on the specifics of the observable degrees of freedom such as the system size. In this case, criticality results from an \emph{extrinsic} effect. Although this mechanism seems to differ from the fine-tuning explanation, they are not entirely unrelated; interacting systems at criticality also generate large fluctuations. In fact, the usual definition of criticality describes not its mechanisms, but rather the behavior of the observable degrees of freedom such as diverging correlation length, scale invariance and nonanalytic thermodynamic functions (see, e.g., Refs.~\cite{landau:80,goldenfeld:92,cardy:96}). Such properties can emerge intrinsically from interactions between components when model parameters are carefully chosen, as is often the case in statistical physics. However, this \emph{intrinsic} mechanism is by no means the only one, nor is it a defining feature of criticality.

Here we introduce a new definition of criticality that spans both intrinsic and extrinsic mechanisms. Using the languages of learning and information theory, we show that a unifying feature of both types of criticality is a divergence of mutual information between two halves of the system.\footnote{Divergent information is a direct result of long-ranged correlations at critical points, at which the entire system becomes correlated. Indeed, it has proved a useful characterization of criticality. See, e.g., Refs.~\cite{grassberger:86,bialek:01a,bialek:01b,tchernookov:13,wilms:12,lau2013information,cohen:15}.} The rates of divergence depend on how the fluctuations scale with the system size and generally differ between intrinsic and extrinsic criticality. We show further that critical systems are equivalent to the problem of learning parameters from \iid\ samples, with the fluctuating fields playing the role of the parameters and system components of \iid\ samples. This learning problem is characterized by diverging information between the parameters and samples. Through the learning-theoretic lens, we interpret intrinsic criticality as a \emph{fractional} learning problem, in which only a fraction of parameters is available for learning due to a sharpening of \emph{a priori} distribution as the system size grows. In contrast, extrinsic criticality has a fixed \emph{a priori} distribution whose entropy, i.e., information available to be learned, does not shrink with the system size.

In addition, we examine the observability of Zipf's law---the inverse relationship between the rank and frequency of system states---commonly used as an empirical signature of criticality~\cite{mora:11,tkacik:15}. While the correspondence between Zipf's law and critical behaviors is expected in the thermodynamic limit~\cite{mora:11}, little is known whether it remains robust under finite sample size and especially in finite systems, for which the definition of criticality becomes blurred. Our analyses reveal that the extrinsic mechanism is a likelier explanation of experimentally observed Zipfian behaviors than typical intrinsic critical systems, such as critical mean-field models, which usually generate fluctuations that are too small. We show further that, when endowed with certain dynamics, intrinsically induced fluctuations can become large enough to support empirically observable Zipf-like distributions over a range of model parameters without the need for fine tuning.

We emphasize that the distinction between both types of criticality is not sharp. Varying the system-size scaling of critical fluctuations leads to a smooth crossover between intrinsic and extrinsic criticality with the latter corresponding to the limit, in which the fluctuations are independent of the system size. Importantly, this crossover behavior means that intrinsically critical fluctuations that depend very weakly on the system size are empirically indistinguishable from extrinsic critical ones.

\section{Information and learning--theoretic view of criticality}

\para{Large fluctuations are a defining feature of criticality.} 
Criticality is often characterized by nonanalytic behaviors of the derivatives of thermodynamic energy or the divergence of correlations of an order parameter~\cite[Ch~XIV]{landau:80}. Such definitions assume the knowledge of the probabilistic model that governs the system, and rely on quantities that are not readily accessible in biology experiments. As a result, they are hard to extend to living systems. Alternatively, one can attempt to define criticality, directly using properties of the joint probability distribution of the system's components which can be estimated directly from data. Zipf's law in the rank-ordered frequencies of the system states is one possibility~\cite{mora:11}. This definition of criticality also describes the samples that encode maximum information about the unknown generative process at a fixed level of compression~\cite{marsili:13,cubero:19,duranthon:21,marsili:22}. More generally, energy is an extensive quantity and thus the central limit theorem ensures that the variance of energy fluctuations for a typical, weakly correlated statistical physics system of size $N$ scales as $\var(E)\!\sim\!N$. A faster growth, $\var(E)\!\sim\!N^\alpha$ with $\alpha\!>\!1$, indicates correlated microscopic fluctuations that violate the \iid\ assumptions of the central limit theorem.\footnote{We assume here that the variance of the energy of each component, $\var(E_i)$, does not depend on $N$. Our argument does not rely on this assumption. If $\var(E_i)\!\sim\!N^\gamma$, then the central limit theorem implies $\var(E)\!\sim\!N^{1+\gamma}$. Then criticality describes the scenarios, in which the energy fluctuations grow faster (or decrease more slowly) than this rate.} We use the presence of these atypically large fluctuations as a \emph{definition} of criticality.

\begin{figure*}
\centering
\includegraphics{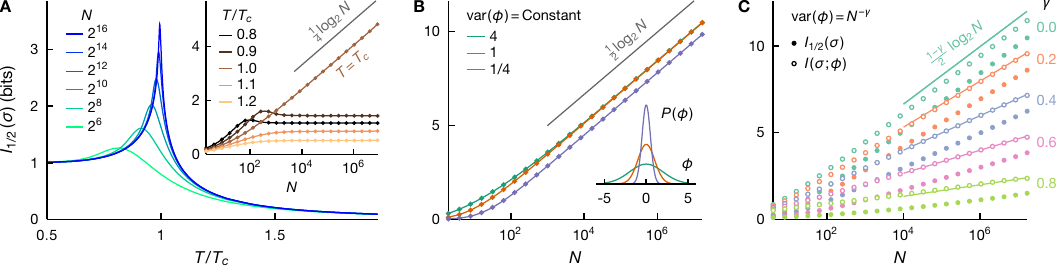}
\caption{\label{fig:info_divergence}%
\textbf{Divergent information signifies criticality.} 
A: Intrinsic criticality in equilibrium generally requires fine tuning. We depict the mutual information between two equal halves of the system, $I_{1/2}(\bm\sigma)\!=\!I(\bm\sigma^A;\bm\sigma^B)$ with $N_A\!=\!N_B\!=\!N/2$, for the fully connected Ising model as a function of temperature for a range of system sizes $N$ (see legend). We see that the information diverges with $N$ only at the critical temperature $T\!=\!T_c$. This divergence is logarithmic, with the asymptotic information given approximately by $\frac{1}{4}\log_2N$ (inset).
B: Extrinsic criticality, on the other hand, emerges without fine tuning. For a system of noninteracting spins, coupled to a common Gaussian fluctuating field $\phi$ (inset), the information $I_{1/2}(\bm\sigma)$ always diverges with $N$. In contrast to the fully connected Ising model at criticality, the asymptotic information grows faster with $N$, $I_{1/2}(\bm\sigma)\approx\frac{1}{2}\log_2N$.
C: The scaling exponent of the critical fluctuation, $\var(\phi)\!\sim\!N^{-\gamma}$, controls the information divergence rate. We illustrate the information between two halves of the system $I_{1/2}(\bm\sigma)$ (filled circles), and between the system and the latent field $I(\bm\sigma;\phi)$ (empty circles) for conditionally independent identical spins, Eq.~\eqref{eq:cond_indep_model}, under a Gaussian fluctuating field $\phi\!\sim\!\mathcal{N}(\mu\!=\!0, s^2\!=\!N^{-\gamma})$ for various values of $\gamma$ (see legend). The asymptotic scaling of the information is in good agreement with the expected logarithmic divergence $\frac{1-\gamma}{2}\log_2N$ (lines), Eq.~\eqref{eq:fractional_learning}. We also see that $I_{1/2}(\bm\sigma)\!\le\!I(\bm\sigma;\phi)$, as expected from the data processing inequality for the Markov chain $\bm\sigma^A$\,--\,$\phi$\,--\,$\bm\sigma^B$.%
}
\end{figure*}

\para{Divergent mutual information between subsystems is another signature of criticality.} 
Large correlated fluctuations allow predictions of the state of one part of the system, $\mathbf{x}^A\!=\!(x_1,x_2,\dots,x_{N_A})$, from measurements of its other part, $\mathbf{x}^B\!=\!(x_{N_A+1},x_{N_A+2},\dots,x_{N_A+N_B})$, with $\mathbf{x}\!=\!(x_1,x_2,\dots,x_{N})$ and $N_A+N_B=N$. Specifically the mutual information between these parts reads~\cite{cover:06}
\begin{equation}
    I(\mathbf{x}^A;\mathbf{x}^B)
    =
    S(\mathbf{x}^A)+S(\mathbf{x}^B)-S(\mathbf{x}),
\end{equation}
where $S(\cdot)$ denotes the entropy of a random variable. At a continuous phase transition, whether equilibrium or not, this mutual information diverges logarithmically,  $I(\mathbf{x}^A;\mathbf{x}^B)\propto\log_2 N$~\cite{grassberger:86,bialek:01a,bialek:01b,tchernookov:13,wilms:12,lau2013information,cohen:15}. We take this divergence as another, \emph{equivalent} definition of criticality. We will show that this definition provides a unifying way of treating extrinsic and intrinsic criticality (see Fig.~\ref{fig:info_divergence}), and to isolate the effects of critical correlations from that due to geometry, we will focus on non-spatial systems here and in the following.

\para{Diverging information also characterizes statistical inference.} 
Generally, the information that \iid\ samples encode about the parameters of a distribution grow logarithmically with sample size $N$ in the limit $N\!\to\!\infty$~\cite{bialek:01a,bialek:01b}. Take, for example, a process that generates \iid\ samples $\mathbf{x}=(x_1,x_2,\dots,x_N)$ from a probability distribution $P(x_i\,|\,\phi)$ where $\phi$ is an unknown, real-valued, and possibly multi-dimensional parameter, which one wants to estimate from $\mathbf{x}$. Given an \emph{a priori} distribution $P(\phi)$, the joint marginal distribution of the samples is
\begin{equation}\label{eq:inference}
    P(\mathbf{x}) 
    = \int\!\!d\phi\,P(\phi)\prod\nolimits_{i=1}^NP(x_i\mid\phi).
\end{equation}
If we divide the samples into two macroscopic groups, i.e., $\mathbf{x}\!=\!(\mathbf{x}^A,\mathbf{x}^B)$, then the mutual information between these groups reads~\cite{bialek:01a} (see also Appendix~\ref{appx:info_divergence})
\begin{equation}
    I(\mathbf{x}^A;\mathbf{x}^B)
    =
    \frac{k}{2}\log_2 \frac{N_AN_B}{N}
    +
    O(N^0)
    \approx
    \frac{k}{2}\log_2 N.
\end{equation}
where $k$ denotes the dimensionality of the parameter $\phi$, or, more precisely, the dimensionality of the subset of the parameters that can be inferred from $\mathbf{x}$.\footnote{We assume that the number of parameters $k$ is much smaller than the sample size $N$. For $1\!\ll\!N\!<\!k$, the information can grow \emph{linearly} with $N$~\cite{ngampruetikorn:22}.} Importantly, the information between the two halves exists only to the extent that both depend on $\phi$. In fact, this information is nearly the same as the information one has about the parameter $\phi$ having observed $\mathbf{x}$, i.e., $I(\mathbf{x}^A;\mathbf{x}^B) \!\approx\! I(\mathbf{x};\phi)$~\cite{bialek:01a} (see Fig.~\ref{fig:info_divergence}C). We emphasize that this logarithmic divergence arises \emph{without} fine tuning and for a broad range of assumptions about $P(\phi)$ and $P(x_i\,|\,\phi)$.

\para{Extrinsic criticality is equivalent to statistical learning.}
We can interpret the information divergence in inference problems as a signature of criticality, imposed extrinsically by the unknown, fluctuating variable $\phi$. 
We consider a physical manifestation of an extrinsically critical system~\cite{schwab:14,aitchison:17} and write down the joint probability of $N$ noninteracting, identical binary spins in an external magnetic field $\phi$,
\begin{equation}
P(\bm\sigma\mid\phi) = \prod\nolimits_{i=1}^N \frac{e^{\phi\sigma_i}}{2\cosh\phi}
\end{equation}
where $\bm\sigma\!=\!(\sigma_1,\sigma_2,\dots,\sigma_N)$ and $\sigma_i\!\in\!\{\pm1\}$. If the field is not constant, but a random variable that fluctuates for different realizations of the system, then marginalizing over this field yields
\begin{equation}\label{eq:cond_indep_model}
P(\bm\sigma) 
= \int\!\! d\phi \, P(\phi) \prod\nolimits_{i=1}^N \frac{e^{\phi\sigma_i}}{2\cosh\phi},
\end{equation}
where $P(\phi)$ is the marginal distribution of the fluctuating field. This equation takes the exact same form as Eq.~\eqref{eq:inference}, with the magnetic field playing the role of the model parameter and the spins of the \iid\ samples. As a result, it follows immediately that $I(\bm\sigma;\phi)\!\approx\!I(\bm\sigma^A;\bm\sigma^B)\!\approx\! \frac{1}{2}\log_2N$~\cite{bialek:01a}, where $\bm\sigma^A$ and $\bm\sigma^B$ denote two halves of the system, see Fig.~\ref{fig:info_divergence}B. In fact, we can derive this result by noticing that the \emph{a posteriori} distribution, $P(\phi\,|\,\bm\sigma)\!\sim\!P(\phi)\exp(\phi\sum_i\sigma_i\!-\!N\ln\cosh\phi)$, sharpens as $N$ grows, with the asymptotic variance decreasing as $\var(\phi\,|\,\bm\sigma)\!\sim\!1/N$. In other words, as our knowledge of the parameter improves with $N$, its \emph{a posteriori} differential entropy decreases as $S(\phi\,|\,\bm\sigma)\!\approx\!-\frac{1}{2}\log_2N$. Given that the \emph{a priori} entropy $S(\phi)$ is finite and does not change with $N$, we obtain 
\begin{equation}\label{eq:logdivergence_ext}
I(\bm\sigma^A;\bm\sigma^B)
\approx
I(\bm\sigma;\phi)
=
S(\phi) - S(\phi\mid\bm\sigma)
\approx
\tfrac{1}{2}\log_2N,
\end{equation}
where the approximations hide the terms of order $O(N^0)$.

We can glean additional insights from a more traditional statistical mechanics argument. First we recast Eq.~\eqref{eq:cond_indep_model} as 
\begin{equation}\label{eq:cond_indep_model2}
P(\bm\sigma) 
=
2^{-N}\int\!\! d\phi \, P(\phi) e^{N(\phi m-\ln\cosh \phi)},
\end{equation}
where $m\!=\!m(\bm\sigma)\!=\!\sum_i\sigma_i/N$ is the magnetization. For large $N$, we evaluate the above integral, using the saddle point approximation and assuming that $P(\phi)$ satisfies certain technical conditions~\cite{schwab:14},
\begin{equation}
P(\bm\sigma) 
\approx
2^{-N} 
\sqrt{\frac{2\pi}{N(1-m^2)}}
P(\phi^*) 
e^{N(\phi^* m-\ln\cosh \phi^*)},
\end{equation}
where $\phi^*\!=\!\phi^*(m)\!=\!\tanh^{-1}m$ denotes the saddle point. Defining the energy function as $E(\bm\sigma)\!=\!-\ln P(\bm\sigma)$ and recalling that the thermodynamic entropy is the logarithm of the density of states---i.e., $\mathcal S(m)\!=\!\ln[\frac{N}{2}\binom{N}{N_+}]$ with $N_+\!=\!N(1+m)/2$---we obtain to the leading order in $N$~\cite{schwab:14}
\begin{equation}\label{eq:energy-entropy-equivalence}
E(m)-\mathcal S(m) =\ln (1-m^2)-\ln P(\phi^*)
+O(N^{-1}),
\end{equation}
which does not grow with $N$. This equivalence between the energy and entropy to all increasing order in $N$ signifies a very strong form of criticality~\cite{mora:11}.

To compute the mutual information, we write down the entropy of the system
\begin{equation}
\!\!
S(\bm\sigma)
=-\sum_{\bm\sigma} P(\bm\sigma)\ln P(\bm\sigma)
=\int_{-1}^{1}\!\!\!\!dm\, e^{\mathcal S(m)-E(m)} E(m).
\!
\end{equation}
Using Eq.~\eqref{eq:energy-entropy-equivalence} and noting that $\frac{d\phi^*}{dm}\!=\!\frac{1}{1-m^2}$, we have
\begin{align}
S(\bm\sigma)
&\approx
\int_{-1}^{1}\!
\frac{dm}{1-m^2} 
P(\phi^*)E(m)
\nonumber\\
&\approx
N\langle\ln2- \phi\tanh\phi+\ln\cosh\phi\rangle_{\phi\sim P(\phi)}
+\tfrac{1}{2}\ln N,
\end{align}
where we drop the terms of order $O(N^0)$ and smaller. As a result, the information between the two halves of the system reads
\begin{equation}
I(\bm\sigma^A;\bm\sigma^B)
\approx
\tfrac{1}{2}\log_2 N_A+\tfrac{1}{2}\log_2 N_B-\tfrac{1}{2}\log_2 N
\approx
\tfrac{1}{2}\log_2 N,
\end{equation}
which exhibits the same logarithmic divergence as one-parameter learning, Eq.~\eqref{eq:logdivergence_ext}.

The specifics of our calculations (conditional independence of spins, binary spins, etc.) are not important for the main result: an extrinsically varying parameter gives the mutual information between two subparts of the system (surface interaction terms excluded) that grows as $\frac{1}{2}\log_2 N$. This is a very specific form of criticality, mathematically equivalent to a problem of learning the said parameter from observations of the system's microstate. For larger dimensional parameters, the calculation generalizes: mutual information is $\frac{1}{2}\log_2 N$ per inferable parameter component~\cite{bialek:01a} (so long as the number of parameters is much smaller than $N$).

\para{Intrinsic criticality is described by `fractional' parameter learning.}
To illustrate the physics of intrinsic criticality, we consider a minimal model of fully connected identical Ising spins, defined by the energy function $E(\bm\sigma)\!=\!-\frac{1}{4N}(\sum_i\sigma_i)^2$. As usual, the joint probability distribution of the spins is $P(\bm\sigma)\!=\!e^{-\beta E(\bm\sigma)}/Z$, where $\beta\!=\!1/T$ denotes the inverse temperature. The partition function $Z$ guarantees proper normalization,
\begin{equation}\label{eq:Z_fullyconnectedising}
Z = \sum_{\bm\sigma}e^{\frac{\beta}{4N}(\sum_i\sigma_i)^2}
=
\sqrt{\frac{N}{\pi\beta}}\int\!\!\! d\phi 
\sum_{\bm\sigma}
e^{-\frac{N}{\beta}\phi^2 +\phi \sum_i\sigma_i},
\end{equation}
where the last equality follows from the Hubbard--Stratonovich transformation. We see that the system consists of fluctuating degrees of freedom that include the spins $\bm\sigma$ as well as a scalar field $\phi$, and the joint distribution of these variables reads
\begin{equation}\label{eq:intrinsic-joint}
P(\bm\sigma,\phi)
=
\frac{1}{Z}
\sqrt{\frac{N}{\pi\beta}}
e^{-\frac{N}{\beta}\phi^2 +\phi \sum_i\sigma_i}. 
\end{equation}
Marginalizing out the spins yields 
\begin{equation}\label{eq:intrinsic-prior}
P_N(\phi)
=
\frac{2^N}{Z}
\sqrt{\frac{N}{\pi\beta}}
e^{-\frac{N}{\beta}\phi^2+N\ln\cosh\phi},
\end{equation}
where the subscript $N$ emphasizes that this distribution is $N$-dependent. Now the distribution over the spin states is
\begin{equation}
P(\bm\sigma)
=
\int\!\!\! d\phi \,
P_N(\phi) \frac{P(\bm\sigma,\phi)}{P_N(\phi)}
=
\int\!\!\! d\phi \,
P_N(\phi)\prod_i \frac{e^{\phi\sigma_i}}{2\cosh\phi}, 
\end{equation}
which is the same form as the extrinsically critical model in Eq.~\eqref{eq:cond_indep_model}, albeit with an \emph{a priori} distribution that depends on $N$.

Intrinsically induced fluctuations behave differently at and away from the critical point. To see this, we recall that $\ln\cosh x\!\approx\!\frac{1}{2}x^2-\frac{1}{12}x^4$ for small $x$. Substituting this approximation into the exponent of Eq.~\eqref{eq:intrinsic-prior} gives, for small $\phi$,
\begin{equation}\label{eq:P_phi_ising}
P_N(\phi)\sim e^{-N(\frac{1}{\beta}-\frac{1}{2})\phi^2-N\frac{1}{12}\phi^4}.
\end{equation}
We see that $\var(\phi)\!\sim\!1/N$ for $\beta\!<\!2$, but $\var(\phi)\!\sim\!1/\sqrt{N}$ at $\beta\!=\!2$. For $\beta\!>\!2$, we have again $\var(\phi)\!\sim\!1/N$ with a prefactor that depends on the curvature around the maxima of $P_N(\phi)$. This change in scaling behaviors is what defines criticality: the fluctuation becomes atypically large only at the critical temperature $\beta_c\!=\!2$, with a variance that scales as $1/\sqrt N$ instead of $1/N$ for this simple fully connected model.

Divergent mutual information emerges only at the critical point, see Fig.~\ref{fig:info_divergence}A. The scaling behaviors of the variance of intrinsically induced fluctuations result in the asymptotic \emph{a priori} differential entropy, $S(\phi)\!\approx\!-\frac{1}{4}\log_2N$ at $\beta\!=\!\beta_c$ and $S(\phi)\!\approx\!-\frac{1}{2}\log_2N$, otherwise. From Eq.~\eqref{eq:intrinsic-joint}, we have $\var(\phi\,|\,\bm\sigma)\!\sim\!1/N$, hence the \emph{a posteriori} differential entropy scales as $S(\phi\,|\,\bm\sigma)\!\approx\!-\frac{1}{2}\log_2N$. We see that the logarithmic divergences in \emph{a priori} and \emph{a posteriori} entropy cancel exactly away from criticality, leading to mutual information that does not grow with $N$. At criticality, on the other hand, we have~\cite{wilms:12}
\begin{equation}\label{eq:logdivergence_fullyconnected}
I(\bm\sigma^A;\bm\sigma^B)\approx I(\bm\sigma;\phi)\approx\tfrac{1}{4}\log_2N.
\end{equation}
We emphasize that the prefactor here is $1/4$ instead of $1/2$---that is, we learn only \emph{half} of a parameter in this intrinsically critical setting.

\para{Fractional learning is typical of intrinsically induced criticality.}
In the above examples of intrinsic and extrinsic criticality, we see that the \emph{a posteriori} differential entropy of the fluctuating field is asymptotically identical, i.e., $S(\phi\,|\,\bm\sigma)\!\approx\!-\frac{1}{2}\log_2N$. The \emph{a priori} entropy, in contrast, differs: $S(\phi)$ does not depend on $N$ for extrinsic fluctuations, whereas $S(\phi)\!\approx\!-\frac{1}{4}\log_2N$ for the fully connected Ising model at criticality. As a result, the mutual information diverges at different rates [Eqs.~\eqref{eq:logdivergence_ext}\,\&\,\eqref{eq:logdivergence_fullyconnected}]. More generally, intrinsically induced fluctuations depend on the system size but need not take the same form as Eq.~\eqref{eq:intrinsic-prior}. For a unimodal fluctuating field with $\var(\phi)\!\sim\!N^{-\gamma}$, the differential entropy reads $S(\phi)\!\approx\!-\frac{\gamma}{2}\log_2N$. Therefore (see Appendix~\ref{appx:info_divergence}),
\begin{equation}\label{eq:fractional_learning}
I(\bm\sigma^A;\bm\sigma^B)\approx I(\bm\sigma;\phi)\approx\frac{1-\gamma}{2}\log_2N.
\end{equation}
Away from criticality, $\var(\phi)\!\sim\!N^{-1}$, and the mutual information is finite: we cannot learn much about the field because, in noncritical thermodynamic systems, fluctuations are small, and we already know almost everything \emph{a priori}. In contrast, critical fluctuations are larger, $\gamma\!<\!1$, so that observing the spins provides information about the specific realization of $\phi$, and hence about the other spins. However, the entropy of intrinsically induced critical fluctuations decreases with $N$ quite generally, $0\!<\!\gamma\!<\!1$, resulting effectively in only a \emph{fraction} of a parameter being available for learning, thereby a decrease in the information from its maximum possible of $\frac{1}{2}\log_2N$. In Fig.~\ref{fig:info_divergence}C, we see that Eq.~\eqref{eq:fractional_learning} agrees well with the asymptotic behavior of mutual information for a range of $\gamma$.

Thus, we have shown that, at least for a simple model, intrinsic criticality can be viewed as a learning problem, where the underlying large fluctuations in the order parameter leave sufficient freedom to learn its specific realization from observations of the system state. The expression of the information in terms of the difference of the \emph{a priori} and the \emph{a posteriori} entropy, Eq.~\eqref{eq:logdivergence_ext}, shows that these results will generalize to other critical systems: critical exponents will govern the \emph{a priori} differential entropy of the order parameter, while the \emph{a posteriori} differential entropy remains approximately $-\frac{1}{2}\log_2N$. Similarly, for multi-dimensional order parameters, each dimension will contribute to the mutual information essentially independently. For correlated fluctuating parameters, the logarithmic divergence rate provides a measure of effective dimensions of the parameters~\cite{bialek:22}. Finally, extrinsic and intrinsic criticality will add up as well so that each extrinsically or intrinsically critical field (i.e., with $N$-independent or $N$-dependent fluctuations) will contribute $1/2$ or a smaller amount to the coefficient in front of $\log_2N$ in the information.

\begin{figure}
\centering
\includegraphics[width=\linewidth]{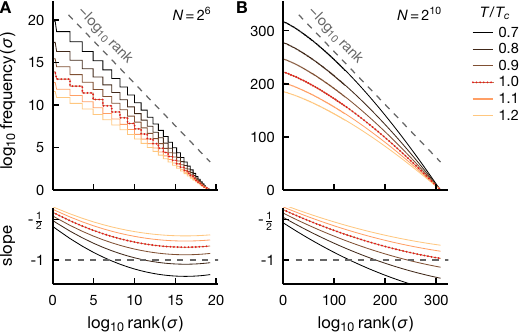}
\caption{\label{fig:exact_rankfreq_ising}%
\textbf{Zipf's law is an inaccurate description of critical mean-field systems.}
We depict exact rank-ordered distributions for the fully connected Ising model [Eq.~\eqref{eq:Z_fullyconnectedising}] at various temperatures $T$ (see legend) for the system size $N\!=\!2^6$ and $2^{10}$ (A and B, respectively). Note that the axes correspond to log-rank and log-frequency. The slope of the smoothed log-log plot illustrates how close the models are to Zipf's law (dashed). Here we obtain the approximate slope (bottom row) by fitting a cubic polynomial to the `knees' of the rank-frequency log-log plot. We see that the rank-frequency plots deviate from the power-law behavior at all temperatures, and this deviation does not appear to improve as the system grows. Importantly the critical model ($T\!=\!T_c$) exhibits Zipf's scaling (dashed) only for the least frequent states in the tail region of the rank-ordered distributions and only when the system is large enough (B). In fact, in a smaller system, the rank-frequency plot can appear more Zipf-like at $T\!<\!T_c$ (A), see also Fig.~\ref{fig:rank1ising}.%
}
\end{figure}

\begin{figure}
\centering
\includegraphics[width=\linewidth]{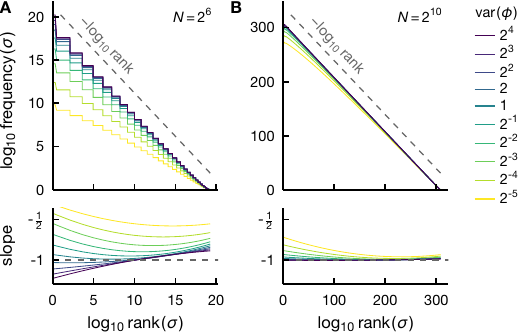}
\caption{\label{fig:exact_rankfreq_gauss}%
\textbf{Zipf's law emerges robustly for adequately large extrinsic fluctuations.}
We show exact rank-ordered distributions for independent spins under a Gaussian fluctuating field of varying variance (see legend). We see that Zipf's law becomes more accurate as the variance of the fluctuations and the system size increase. However, the rank-ordered distributions approach Zipf's scaling over the entire range only when the system is adequately large (B). Here we obtain the approximate slope (bottom row) using the same method as in Fig.~\ref{fig:exact_rankfreq_ising}.%
}
\end{figure}

\section{Signature of criticality in finite systems}

In reality, we can only observe a finite number of components, and the analysis of asymptotic behaviors, while instructive, becomes less precise. To this end, we now turn to the observability of criticality in finite systems. Criticality admits a number of potentially observable signatures. We focus on the properties of the empirical joint distribution of the system components, which can be constructed directly from observational data and thus is readily usable in the context of living systems. A critical system is expected to exhibit Zipf's law, i.e., an inverse relationship between ranks and frequencies of the system states~\cite{mora:11}.

\begin{figure*}
\centering
\includegraphics[width=\linewidth]{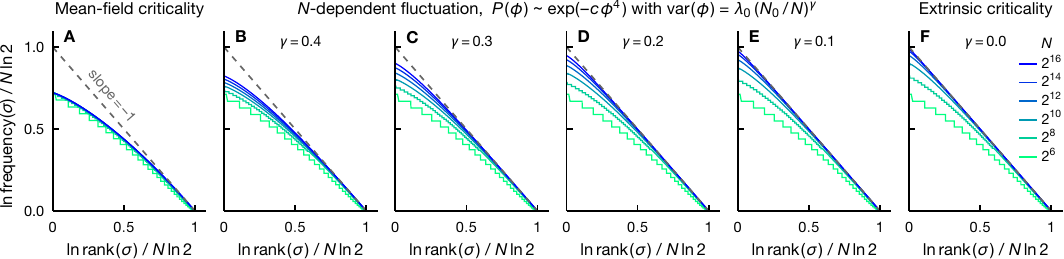}
\caption{\label{fig:exact_rankfreq}%
\textbf{Rank-ordered distributions exhibit a smooth crossover between intrinsic and extrinsic criticality.}
Normalized exact rank-frequency plots for a range of system sizes $N$ (see legend in F) illustrate a gradual crossover from the fully connected Ising model at criticality (A) to spins under an extrinsic fluctuating field (F). This crossover is induced by varying the scaling exponent of the variance of fluctuations $\var(\phi)\!\sim\!N^{-\gamma}$, from $\gamma\!=\!1/2$ for the mean-field criticality in A [see Eq.~\eqref{eq:intrinsic-prior}] to $\gamma\!<\!1/2$ for non-mean-field intrinsic criticality in B-E (see label) and $\gamma\!=\!0$ for extrinsic criticality in F.
We see that the rank-frequency plots approach Zipf's law (dashed) as the system grows, and at a faster rate for a smaller scaling exponent $\gamma$. For mean-field criticality, the agreement with Zipf scaling neither improves, nor degrades, with increasing $N$ (Panel A). 
In B-E, the prior over fluctuations takes the form $P(\phi)\!\sim\!e^{-c\phi^4}$ where $c\!\propto\!N^{2\gamma}$ [cf.\ Eq.~\eqref{eq:P_phi_ising}]. In all panels, the fluctuation variance at $N\!=\!64$ (the smallest $N$ shown) is the same and equal to that of the critical mean-field case. That is, in B-F, we have $\var(\phi)\!=\!\lambda_0(64/N)^\gamma$ with $\lambda_0$ denoting the variance of critical mean-field fluctuation at $N\!=\!64$.%
}
\end{figure*}

First we recap how Zipf's behavior emerges from large fluctuations~\cite{schwab:14,aitchison:17} in the asymptotic limit. Consider the joint distribution of $N$ conditionally independent spins, 
\begin{equation}\label{eq:P_sigma_int}
P(\bm\sigma)=\int\!\!\!d\phi\,P(\phi)\prod\nolimits_{i=1}^N P(\sigma_i\mid\phi).
\end{equation}
We see that $\sum_i\ln P(\sigma_i\,|\,\phi)\!\sim\!N$, and thus $P(\bm\sigma\,|\,\phi)$ becomes a sharper function of $\phi$ as $N$ increases, with a characteristic width that scales as $1/\sqrt{N}$. This scaling sets a threshold above which fluctuations are critical---that is, critical fluctuations are characterized by $\var(\phi)\!\sim\!N^{-\gamma}$ with $\gamma\!<\!1$. As $N\!\to\!\infty$, a critical prior, $P(\phi)$, appears flat with respect to $P(\bm\sigma\,|\,\phi)$ which becomes infinitely sharp. Therefore, we can make the approximation,
\begin{equation}
P(\bm\sigma\mid\phi)\approx e^{-\mathcal{S}(\phi)}\delta(\phi-\phi^*_{\bm\sigma}),
\end{equation}
where $\mathcal{S}(\phi)\!\equiv\!\ln\sum_{\bm\sigma}\delta(\phi-\phi^*_{\bm\sigma})$ plays the role of the thermodynamic entropy and $\phi^*_{\bm\sigma}$ is the maximum of $P(\bm\sigma\,|\,\phi)$, assuming only one exists. Substituting the above approximation into Eq.~\eqref{eq:P_sigma_int} yields
\begin{equation}
    P(\bm\sigma)\approx 
    P(\phi^*_{\bm\sigma})e^{-\mathcal{S}(\phi^*_{\bm\sigma})}
    \equiv
    e^{-\mathcal{E}(\phi^*_{\bm\sigma})},
\end{equation}
which illustrates that the energy function depends on $\bm\sigma$ only through $\phi^*_{\bm\sigma}$, i.e., $E(\bm\sigma)\!=\!-\ln P(\bm\sigma)\!=\!\mathcal{E}(\phi^*_{\bm\sigma})$. Importantly, the above equation signifies Zipf's law via the equivalence between the extensive parts of the entropy and energy~\cite{mora:11},
\begin{equation}
    \lim_{N\to\infty}\left(\mathcal{E}(\phi)-\mathcal{S}(\phi)\right)/N 
    = \lim_{N\to\infty}-N^{-1}\ln P(\phi)
    = 0.
\end{equation}
For mean-field criticality, $\ln P(\phi)\!\sim\!N\phi^4$ [Eq.~\eqref{eq:P_phi_ising}] and the above cancellation holds when $\phi$ is adequately small. This condition is guaranteed for a typical realization of $\phi$ since $\var(\phi)\!\sim\!1/\sqrt{N}$. For extrinsic criticality, $\ln P(\phi)\!\sim\!O(N^{0})$ and the entropy-energy equivalence needs not rely on $\phi$ being small.

\begin{figure}
\centering
\includegraphics[width=\linewidth]{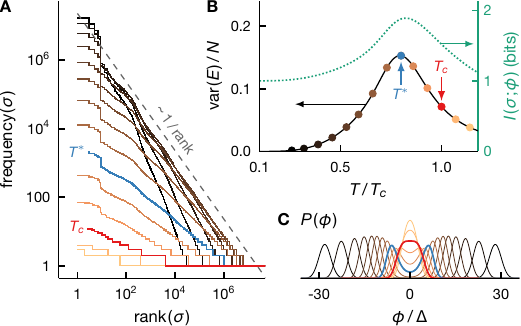}
\caption{\label{fig:rank1ising}%
\textbf{Empirical Zipf behavior needs not coincide with the critical point of the system.} 
A: We depict empirical rank-ordered distributions of the rank-one Ising model [Eqs.~(\ref{eq:rank1_ising_energy}-\ref{eq:rank1_ising_prior})] at various temperatures (see color legend in B). We see that the distribution is closest to Zipf's law at an intermediate temperature, significantly lower than the critical temperature $T_c$. 
As an empirical, alternative definition of a critical temperature for finite systems, we consider the maximum of the specific heat or, equivalently the energy variance. This definition results in $T^*\!<\!T_c$ (see B), which is still significantly higher than the temperature that exhibits approximate Zipf behavior. 
B: The energy variance $\var(E)$ (solid, left axis) and mutual information $I(\bm\sigma;\phi)$ (dashed, right axis) provide measures of correlations in the system. Both maximize below $T_c$ roughly at the same temperature ($T^*$ for $\var(E)$). 
C: The prior $P(\phi)$ changes from unimodal to bimodal at $T_c$, which indicates maximum correlations in the thermodynamic limit. However, in finite systems, $P(\phi)$ cannot be accurately described by its behavior near maxima. We see that the specific heat peaks at $T^*\!<\!T_c$ (see B), where $P(\phi)$ is bimodal but with significant density at $\phi\!=\!0$.
Here the results are for a system of $60$ spins, $10^8$ realizations per model and $w_i\!\sim\!\mathcal{N}(\mu\!=\!1,s\!=\!0.3)$ [see Eq.~\eqref{eq:P_sigma_phi}].%
}
\end{figure}

In Fig.~\ref{fig:exact_rankfreq_ising}, we depict exact (infinite samples) rank-frequency plots for the fully connected Ising model at a range of temperatures. We see a clear deviation from the power-law behavior at all temperatures. The rank-frequency plot at $T_c$ approaches Zipf's scaling, but only in the tail region and for an adequately large system (Fig.~\ref{fig:exact_rankfreq_ising}B). In a smaller system, Zipf's law can appear more accurate at $T\!<\!T_c$ (Fig.~\ref{fig:exact_rankfreq_ising}A). This disconnect between the Zipf behavior and criticality in mean-field models is likely to be more visible under finite samples, which can only probe parts of the exact rank-frequency plots (see also Fig.~\ref{fig:rank1ising}). We emphasize here that the number of observations required to resolve the tail of this rank-frequency plot would be experimentally impractical, $2^N\!\sim\!10^{20}$ for $N\!=\!2^6$ and $2^N\!\sim\!10^{300}$ for $N\!=\!2^{10}$ (Fig.~\ref{fig:exact_rankfreq_ising}A\&B).

On the other hand, Fig.~\ref{fig:exact_rankfreq_gauss} shows that the rank-ordered distributions of identical spins under extrinsic fluctuations become more Zipf-like over the \emph{entire} range of frequencies and ranks as the fluctuation variance increases and as the system grows. However, the rank-frequency plots approach Zipf's law only when the system is sufficiently large (Fig.~\ref{fig:exact_rankfreq_gauss}B).

Although mean-field criticality results in a Zipf-like rank-frequency plot only in the hard-to-observe tail region, intrinsic criticality with a more general critical exponent---i.e., $\var(\phi)\!\sim\!N^{-\gamma}$ with $\gamma\!\ne\!1/2$---can generate rank-ordered distributions that are much closer to Zipf's law. In Fig.~\ref{fig:exact_rankfreq}, we compare the system-size dependence of the rank-frequency plots for the fully connected Ising model at $T_c$ ($\gamma\!=\!1/2$) to several models with larger fluctuations, including those potentially induced by intrinsic criticality of non-mean-field models ($0\!<\!\gamma\!<\!1/2$) as well as the limiting case of extrinsic criticality ($\gamma\!=\!0$). Figure~\ref{fig:exact_rankfreq}A shows that the critical mean-field model exhibits Zipf scaling only in the tail region of the rank-ordered distribution (see also Fig.~\ref{fig:exact_rankfreq_ising}) and the agreement with Zipf's law does not improve, nor degrade, as the system grows. On the other hand, we see that if the fluctuation variance decreases more slowly with $N$, i.e., $\gamma\!<\!1/2$, Zipf's law gradually becomes more accurate as $N$ increases, see Fig.~\ref{fig:exact_rankfreq}B-F. This behavior implies that empirically observed Zipf behavior is likely to indicate either extrinsic criticality or intrinsic criticality of non-mean-field type, characterized by critical fluctuations that scale only weakly with the system size.

\begin{figure*}
\centering
\includegraphics[width=\textwidth]{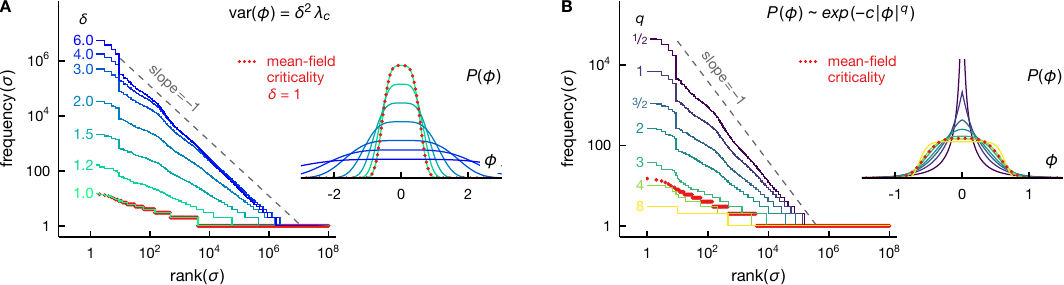}
\caption{\label{fig:finitesampleszipf}%
\textbf{Critical mean-field fluctuations are too small to support empirically observable Zipf behavior.}
Panels A and B illustrate the effects of the width and the structure of the tails of the \emph{a priori} distribution $P(\phi)$ (inset) on finite-sample rank-frequency plots, respectively. The red dotted curves correspond to the rank-one Ising model at criticality [Eqs.~(\ref{eq:rank1_ising_energy}-\ref{eq:rank1_ising_prior})]. 
In A, we consider linear scaling of critical mean-field fluctuations such that $\var(\phi)\!=\!\delta^2\lambda_c$, where $\lambda_c$ is the fluctuation variance of the rank-one Ising model at $T_c$, for various scaling coefficient $\delta$ (see legend). We see that critical mean-field fluctuations ($\delta\!=\!1$) do not result in Zipf behavior. Increasing the fluctuation variance while maintaining the overall shape of the fluctuation prior produces rank-frequency plots that progressively appear closer to Zipf scaling (dashed). 
In B, we consider the fluctuating field of the form $P(\phi)\!\sim\!e^{-c|\phi|^q}$. We vary the probability in the tails of $P(\phi)$ with the shape parameter $q$ (see legend), and choose the scale parameter $c$ such that the fluctuation variance is fixed and equal to that of critical mean-field fluctuations (red dotted lines). Decreasing $q$ increases the probability of large $\phi$, i.e., puts more mass in the tails of $P(\phi)$, and improves the agreement between rank-frequency plots and Zipf's law. 
Overall, while instructive in understanding critical behaviors, mean-field models are an unlikely candidate for explaining experimentally observed Zipf behavior. 
Here the results are for a system of $60$ spins, $10^8$ realizations per model and $w_i\!\sim\!\mathcal{N}(\mu\!=\!1,s\!=\!0.3)$ [see Eq.~\eqref{eq:P_sigma_phi}].%
}
\end{figure*}

\section{Zipf's law as a signature of criticality under finite samples}

Experimental measurements are finite in not only the number of observable degrees of freedom but also the number of observations. We now turn to examine the behavior of rank-frequency plots constructed from finite samples.

In the following, we generalize our conditionally independent model, Eq.~\eqref{eq:cond_indep_model}, to describe nonidentical spins,
\begin{equation}\label{eq:P_sigma_phi}
P(\bm\sigma\mid\phi)=\prod_i \frac{e^{\sigma_iw_i\phi}}{2\cosh w_i\phi}.
\end{equation}
Here $w_i\!\sim\!O(N^0)$ is the coupling strength between the spin $\sigma_i$ and the field $\phi$, which can differ from one spin to another. For convenience, we also define 
\begin{equation}\label{eq:Delta}
\Delta \equiv 1/\sqrt{\sum\nolimits_iw_i^2}, 
\end{equation}
which is the characteristic width of $P(\bm\sigma\,|\,\phi)$ at large $N$, i.e., $P(\bm\sigma\,|\,\phi)\!\sim\!e^{-(\phi-\phi^*_{\bm\sigma})^2/2\Delta^2}$ with $\phi^*_{\bm\sigma}\!=\!\sum_iw_i\sigma_i/\sum_iw_i^2$. Following the argument in the preceding section, we expect Zipf behavior when $\var(\phi)\!\gg\!\Delta^2\!\sim\!O(N^{-1})$.

Similarly, for mean-field criticality, we consider the rank-one Ising model, which generalizes the fully connected Ising model to nonidentical spins and is defined by the energy function,
\begin{equation}\label{eq:rank1_ising_energy}
E(\bm\sigma)=-\frac{1}{4N}\sum\nolimits_{ij}w_iw_j\sigma_i\sigma_j=-\frac{1}{4N}\left(\sum\nolimits_iw_i\sigma_i\right)^2,
\end{equation}
where $w_i\!\sim\!O(N^0)$ and $w_iw_j$ describes the pairwise interaction between spins $i$ and $j$. This model can be recast as a conditionally independent model, with the conditional distribution of the spins given by Eq.~\eqref{eq:P_sigma_phi} and an \emph{a priori} distribution that depends on both $N$ and $\{w_i\}$ (see Appendix~\ref{appx:rank1}),
\begin{equation}\label{eq:rank1_ising_prior}
    P_N(\phi)
    =
    \frac{2^N}{Z}
    \sqrt{\frac{N}{\pi\beta}}
    e^{-\frac{N}{\beta}\phi^2 + \sum_i \ln\cosh w_i\phi},
\end{equation}
where $Z\!=\!\sum_{\bm\sigma}e^{-\beta E(\bm\sigma)}$. The thermodynamic critical temperature, $\beta_c\!=\!2N/\sum_iw_i^2$, marks the point at which this distribution changes from unimodal to bimodal.

Figure~\ref{fig:finitesampleszipf} illustrates that critical mean-field fluctuations are too small to generate experimentally observable Zipf's law. We consider a system of 60 conditionally independent spins under a number of \emph{a priori} distributions, $P(\phi)$ (see inset), including that induced by a rank-one Ising model at $T_c$ [Eq.~\eqref{eq:rank1_ising_prior}]. For each \emph{a priori} distribution, we draw $10^8$ \iid\ realizations of the system and construct an empirical rank-frequency plot. In Fig.~\ref{fig:finitesampleszipf}A, we see that the rank-one Ising model at $T_c$ does not produce Zipf's law. Yet, if we make the fluctuation larger while fixing the shape (standardized moments) of the fluctuation prior, the resulting rank-frequency plot edges closer to Zipf scaling. However, the fluctuation variance is not the only factor that controls the behavior of the rank-frequency plot. In Fig.~\ref{fig:finitesampleszipf}B, we see that at a fixed variance, an \emph{a priori} distribution with thicker tails (larger standardized moments) produces a more Zipf-like rank-order plot. We emphasize that the resolution of these plots, especially in the tails, is limited by the number of samples. While we do not rule out the possibility that mean-field criticality may exhibit Zipf behavior in the tail region (see also Fig.~\ref{fig:exact_rankfreq_ising}), observing such behavior would require orders of magnitude more samples than $10^8$ and would therefore be experimentally impractical.

Extrapolating the asymptotic critical temperature to finite systems is, of course, somewhat dubious. In Fig.~\ref{fig:rank1ising}, we consider another frequently used empirical definition of criticality which identifies the critical point with the maximum in the specific heat or equivalently the energy variance. In the asymptotic limit $N\!\to\!\infty$, this definition is identical to the thermodynamic critical temperature $T_c$. For finite systems, however, the specific heat maximum occurs at a lower temperature $T^*\!<\!T_c$ (Fig.~\ref{fig:rank1ising}B). This temperature also coincides roughly with another possible empirical definition of criticality, namely the maximum of the mutual information $I(\bm\sigma;\phi)$ which indicates maximum correlations and learnability (Fig.~\ref{fig:rank1ising}B). In Fig.~\ref{fig:rank1ising}A, we see that lowering the temperature of the rank-one Ising model from $T_c$ to $T^*$ makes the system closer to, but still visibly different from, Zipf's law. In fact, the closest agreement to Zipf's law occurs at an even lower temperature. Two factors contribute to this intriguing temperature dependence. First, for mean-field criticality, Zipf scaling is expected only in the tail of the rank-frequency plot (see Fig.~\ref{fig:exact_rankfreq_ising}B), which requires a very large number of samples to resolve. Second, the correspondence between criticality and Zipf behavior is blurred in finite systems with the tendency for Zipf's law to be more accurate at subcritical temperatures (see Fig.~\ref{fig:exact_rankfreq_ising}A). In sum, we demonstrate that for mean-field models, empirically observable Zipf behavior can be completely uncoupled from the usual notion of criticality.

Figure~\ref{fig:rank1ising}C illustrates the interaction-induced \emph{a priori} distribution $P(\phi)$ at various temperatures. We see that $P(\phi)$ is flat around its maximum at $T_c$. In the thermodynamic limit, this condition leads to non-Gaussian fluctuations which break the central limit theorem and generate critical correlations. However, in finite systems, the field $\phi$ is not well described by fluctuations in the immediate vicinity of its most likely values. We see that at the specific heat maximum $T^*$, the \emph{a priori} distribution has a non-negligible density at $\phi\!=\!0$ even though it is bimodal with maxima at $\phi\!\ne\!0$. This distribution results in a larger fluctuation than at $T_c$, resulting in higher energy variance as well as a rank-ordered plot closer to Zipf's law. As the temperature drops below $T^*$, the most likely field values move further away from zero. Larger values of $\phi$ suppress the variability of the system: each spin $\sigma_i$ aligns with $\mathrm{sign}(w_i\phi)$ with increasing probability, thereby the decrease in the energy variance. At high temperatures $T\!>\!T_c$, the \emph{a priori} distribution becomes sharply peaked at $\phi\!=\!0$, resulting in more random systems and thus a decrease in correlations. Rank-ordered plots also reflect this competition; reduced variability leads to a rank-ordered plot that decays faster than Zipf's law at low $T$, whereas increased randomness yields a plot that appears flatter than Zipf's law at high $T$.

\begin{figure}
\centering
\includegraphics[width=\linewidth]{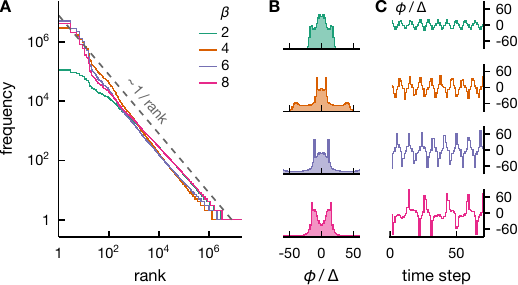}
\caption{\label{fig:nofinetuning}%
\textbf{Zipf's law can emerge from intrinsically induced fluctuations without fine tuning.}
A: Rank-ordered distributions for a dynamical spin model, Eq.~\eqref{eq:dynamical_model}, display Zipf's law for a range of effective inverse temperatures $\beta$ (see legend). 
B\&C: We illustrate typical empirical distributions of the fluctuating field $\phi$ and its typical dynamics (same legend as A).
The results shown are for a system of $60$ spins, $10^8$ realizations per model and $w_i\!\sim\!\mathcal{N}(\mu\!=\!1,s\!=\!0.6)$, and we set $\alpha\!=\!-0.8$ [see Eq.~\eqref{eq:dynamical_model}].%
}
\end{figure}

\para{Intrinsic fluctuations can lead to Zipf's law without fine tuning.} 
So far we see that, in the absence of external fluctuations, the empirical signatures of large correlated fluctuations, such as Zipf-like distributions, are hard to observe. While this statement is true for equilibrium systems, critically large intrinsic fluctuations can emerge generically when the system is endowed with certain dynamics. To illustrate this point, we consider a discrete-time, dynamical generalization of the conditionally independent spin model [Eq.~\eqref{eq:P_sigma_phi}], 
\begin{equation}\label{eq:dynamical_model}
P(\bm\sigma_{t+1}\mid\phi_t)
=
\prod_i \frac{e^{\sigma_{i,t}w_i\phi_t}}{2\cosh w_i\phi_t},
\end{equation}
with $\phi_t\!=\!\beta(m_t+\alpha m_{t-1})$ and $m_t\!=\!\frac{1}{N}\sum_i w_i\sigma_{i,t}$. Here the index $i$ labels each spin and $t$ the time step. The parameter $\beta$ controls the stochasticity of the spins much like the inverse temperature and $\alpha$ couples states separated by two time steps. We illustrate the dynamics of the fluctuating field $\phi_t$ for various $\beta$ and the corresponding \emph{a priori} distributions in Fig.~\ref{fig:nofinetuning}B\&C, respectively. In Fig.~\ref{fig:nofinetuning}A, we see that this model can result in Zipf behavior over a range of model parameters, demonstrating that fine tuning is not a requirement for empirically observable Zipf behavior in the absence of external fields. Although the fluctuations are generated entirely internally, we can interpret this emergence of Zipf's law as extrinsic criticality. To see this, we note that the model parameters $\alpha$ and $\beta$ control the amplitude of the oscillations and thus the fluctuation variance (Fig.~\ref{fig:nofinetuning}B\&C). The system size $N$ only determines the stochasticity of the dynamics. As a result, the \emph{a priori} distribution depends only weakly on $N$ and becomes completely independent of $N$ in the asymptotic limit. This diminishing system-size dependence makes the resulting fluctuations indistinguishable from extrinsic ones.

\section{Discussion}

Here we introduced general definitions of criticality encompassing extrinsically and intrinsically critical systems, and we showed that information-theoretic and learning-theoretic considerations allow us to view all non-spatial, critical systems on a similar footing. Namely, criticality leads to the logarithmic divergence in the information between two subsystems or between the system's observable degrees of freedom and its fluctuating latent field. The coefficient in front of the divergence is semi-integer for systems with extrinsic criticality, and other fractions for intrinsic criticality. Both situations can be viewed as learning the parameter from measurements of the state of system components, and the \emph{a priori} variance of this parameter is independent of the system size (extrinsic criticality) or decreases as the system size grows (intrinsic criticality).

We focused on the scenario where the most critical a system can be is when it has $ \frac{1}{2}\log_2N$ bits of mutual information per dimension of the order parameter, which is equivalent to the \iid\ learning problems. Additional intrinsic couplings would then reduce the \emph{a priori} variance of the order parameter and hence reduce the mutual information. However, some biological systems may have the \emph{a priori} parameter variance that increases with $N$~\cite{humplik:17}, so that more than $\frac{1}{2}\log_2N$ bits are contributed to the mutual information per latent field.

One can imagine this happening in an optimally designed sensory system, where spins are coupled to the field in a way to reduce the redundancy of the information they obtain about it. Investigating the properties of such systems from information-theoretic, learning, and statistical physics angles is clearly needed. Similarly, it is worth investigating systems in which the mutual information between macroscopic parts scales as a sublinear power of $N$, (rather than a logarithm), which correspond to an infinite number of latent fields with hierarchically smaller \emph{a priori} variances~\cite{bialek:01a}.\footnote{We note that some models, such as the random energy model, can generate extensive information~\cite{ngampruetikorn:23random}, but these models are generally not learnable from finite data and hence of less experimental relevance~\cite{bialek:20}.} Finally, since all of our learning and information-theoretic arguments are asymptotic, and $O(1)$ corrections may not be negligibly small compared to $\log_2 N$, subleading corrections are also worth investigating.

In addition, there is a striking similarity between the critical behavior of mutual information in classical systems and entanglement entropy, a quantum information-theoretic measure of correlations. At quantum critical points, long-ranged correlations lead to diverging entanglement entropy, violating the \emph{area law}~\cite{eisert:10}. For infinite quantum critical spin chains, this divergence is logarithmic in the subsystem size with a universal prefactor that is related to the central charge of the corresponding conformal field theory~\cite{holzhey:94,vidal:03,calabrese:04,refael:04,ryu:06}. It would be interesting to develop a learning-theoretic picture of quantum criticality and explore whether and how the central charge relates to the effective number of latent parameters.

We also investigated the observability of an empirical signature of criticality, namely Zipf's law. While the correspondence between Zipf behavior and criticality is precise in the thermodynamic limit, whether it holds for a finite system depends on how critical the system is, i.e., how large the \emph{a priori} variance $\var(\phi)$ is, compared to the width of the conditional distribution $P(\bm\sigma\,|\,\phi)$. For extrinsic criticality, Zipf's law emerges robustly under an adequately broad \emph{a priori} fluctuation distribution. Intrinsic criticality, on the other hand, does not always induce large enough fluctuations to support Zipf behavior. In particular, mean-field critical fluctuations are too small to generate Zipf's law even in the infinite-sample limit. Indeed, under finite samples, the closest agreement to Zipf's law can occur at a temperature significantly lower than the thermodynamic critical temperature as well as the specific heat maximum, an alternative, empirical definition of a critical point. This approximate Zipf behavior at intermediate temperature results from the competition between order-promoting interactions and thermal noise, and is unlikely to be a signature of equilibrium intrinsic criticality in the usual sense.

Perhaps the disconnect between intrinsic criticality and Zipf behavior in finite systems is unsurprising, not least because of the blurred notion of criticality away from the thermodynamic limit. While using the specific heat maximum to indicate criticality is tempting, it requires assumptions on the probabilistic model that describes the data. Zipf behavior offers an alternative, model-free definition, but as we showed, it can be nontrivial to observe in intrinsically critical systems. We emphasize that no finite-system definition of criticality captures all of its thermodynamic signatures; for instance, finite-sample Zipf behavior does not correspond to maximum energy fluctuations or maximum correlations (see Fig.~\ref{fig:rank1ising}).

While our asymptotic analysis suggests that the correspondence between intrinsic criticality and Zipf behavior is more precise for larger systems, we focus on a relatively small system of 60 spins (Figs.~\ref{fig:finitesampleszipf}-\ref{fig:nofinetuning}) since it is more relevant to real measurements. In particular, a well-sampled rank-ordered plot becomes exceedingly difficult to achieve as the system size grows. For example, a rank-ordered distribution for the critical rank-one Ising model with 80 spins shows almost no structure even at $10^8$ samples (see Fig.~\ref{fig:subsampling}). This loss of structure due to finite samples is less severe for extrinsic fluctuations but the agreement with Zipf's law degrades with increasing $N$ (Fig.~\ref{fig:N80_extrinsic}), in contrast to the infinite-sample case, in which Zipf's law becomes more accurate as the system grows (Fig.~\ref{fig:exact_rankfreq}F).

We show further that some oscillatory systems can generate observable Zipf's law without fine tuning or external fluctuations. We argue that the mechanism behind this behavior is mathematically equivalent to the extrinsic mechanism since the scale of the collective dynamics---hence the variance of the fluctuations once the time variable is integrated out---is often independent of the system size. Indeed, collective oscillations are common both in mathematical models (e.g., Refs.~\cite{mirollo:90,acebron:05}) and in biological systems (e.g., Refs.~\cite{buzsaki:04,gregor:10,ditalia:22}). Our results suggest that models with a global dynamical variable that stays within a certain range could offer another plausible explanation for empirically observed Zipf's law, without the need for extrinsic fluctuations.

Finally, we discuss how subsampling may affect the observability of Zipf behavior. Many experiments do not measure the system in its entirety and what we can vary is the number of the observed components rather than the system size. Perhaps the unobserved degrees of freedom could play the role of an extrinsic source of fluctuations for the observed ones, resulting in extrinsically induced criticality and thus making observation of Zipf's law more probable. However, our simple model of intrinsic criticality does not support this thinking. Suppose we observe $K$ out of the total of $N$ spins. Intrinsic fluctuations quite generally become smaller with $N$, i.e., $\var(\phi)\!\sim\!N^{-\gamma}$ with $0\!<\!\gamma$, whereas the width of the conditional probability $P(\bm\sigma\,|\,\phi)$ decreases with $K$, i.e., $\Delta\!\sim\!K^{-1/2}$ [cf.\ Eq.~\eqref{eq:Delta}]. As a result, the relative fluctuation variance is $\var(\phi)/\Delta^2\!\sim\! (K/N) \times N^{1-\gamma}$ with $\gamma\!<\!1$ for critical systems. In other words, decreasing the observable fraction makes the fluctuation appear smaller and Zipf's law less likely (even though the smaller number of degrees of freedom makes it easier to obtain better-sampled rank-frequency plots, see Fig.~\ref{fig:subsampling}). This effect is even more acute when the system size far outnumbers the observed components, e.g., a recording of neural spikes in the brain.

Real systems can of course be more complicated than our simple model. For example, in spatially extended systems, the order parameter could be a field in space and the number of inferable parameters, e.g., the Fourier components of the order parameter, can depend on how many spins we observe. Thus, the bits available to be learned can depend on the number of observed spins, even though the \emph{a priori} variances of the parameters do not (as they are set by the size of the entire system). Investigations of criticality and its empirical signatures in this setting are in order.

We end by pointing out that many biological critical systems become more Zipf-like as they grow~\cite{mora:11}, which begs the question of why this happens. As pointed out in Ref.~\cite{schwab:14}, consider a sensory system that is learning the state of the outside world (that is, responds to its different values differently); one would expect this system to be constructed in a way not to decrease the variability of the world when the system size grows. Such systems would always be critical, and specifically extrinsic critical, maybe explaining their ubiquity.

\begin{acknowledgments}
We thank William Bialek, Stephanie Palmer, and Pankaj Mehta for valuable discussions. VN and DJS acknowledge support from the National Science Foundation, through the Center for the Physics of Biological Function (PHY-1734030). DJS is supported in part by the Simons Foundation and by the Sloan Foundation. IN was supported in part by the Simons Foundation Investigator grant and the NIH grants 1R01NS099375 and 2R01NS084844.
\end{acknowledgments}

\appendix

\begin{figure}
\centering
\includegraphics[width=\linewidth]{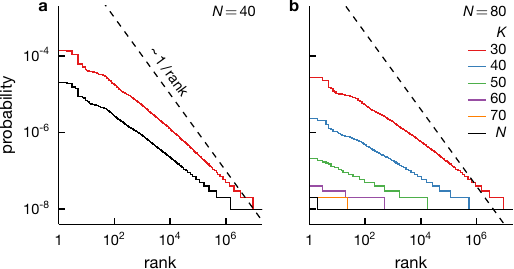}
\caption{\label{fig:subsampling}%
\textbf{Subsampling does not make intrinsically critical systems appear more Zipf-like.}
We depict the rank-ordered frequency for the rank-one Ising model of $N$ spins, Eq.~\eqref{eq:rank1_ising_energy}, at the critical temperature, for $N\!=\!40,80$ (a\,\&\,b), and a range of observed subsystem size $K$ (see legend). We see that limiting observation to a fraction of the full system, i.e., $K\!<\!N$, does not result in more Zipf-like behavior (dashed). It leads however to more structured rank-ordered distributions, especially when the system is large (b), since a system of fewer spins requires a smaller sample size to be well-sampled. The results shown are for $10^8$ realizations per model and $w_i\!\sim\!\mathcal{N}(\mu\!=\!1,s\!=\!0.3)$.%
}
\end{figure}

\begin{figure}
\centering
\includegraphics[width=\linewidth]{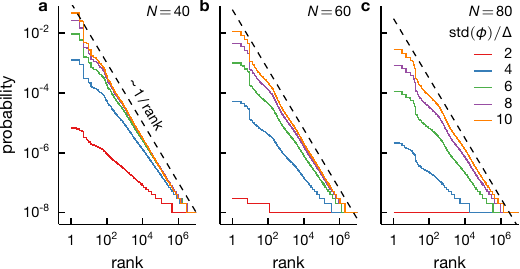}
\caption{\label{fig:N80_extrinsic}%
\textbf{Zipf's law emerges robustly from large extrinsic fluctuations.} 
We display rank-ordered distributions for the conditionally independent model with $N$ spins for $N\!=\!40,60,80$ (left to right) under a Gaussian fluctuating field of different widths (see legend). When the fluctuation $\phi$ is large compared to the width of the conditional distribution $P(\bm\sigma\,|\,\phi)$---i.e., when $\std(\phi)\!\gg\!\Delta$, see Eq.~\eqref{eq:Delta}---the rank-ordered plots exhibit Zipf behavior (dashed). The empirical rank-ordered distribution displays meaningful structures only when constructed from adequate samples. Larger systems require more samples; for $N\!=\!80$, the distribution is completely flat for $\std(\phi)/\Delta\!=\!2$ even with $10^8$ samples (c). The results shown are for $10^8$ realizations per model and $w_i\!\sim\!\mathcal{N}(\mu\!=\!1,s\!=\!0.3)$.%
}
\end{figure}

\section{Mutual information in conditionally independent models\label{appx:info_divergence}}

In this appendix, we consider conditionally independent spin models and derive the leading contribution to the mutual information between the spins and the fluctuating field and between two halves of the system in the many-spin limit.

First, we write down the probability distribution of the spins,
\begin{equation}\label{eq:condindepmodel}
    P(\bm\sigma)
    =
    \int\!\!\!d\phi\, P(\phi)
    \prod\nolimits_iP(\sigma_i\mid\phi),
\end{equation}
where $P(\phi)$ denotes the distribution of the fluctuating field $\phi$. The conditional probability of each spin reads 
\begin{equation}
P(\sigma_i\mid\phi)
=
\frac{e^{\sigma_iw_i\phi}}{2\cosh w_i\phi},
\end{equation}
where $w_i$ parametrizes the influence of the fluctuating field $\phi$ on spin $i$. For convenience, we introduce
\begin{equation}
g(\phi;\bm\sigma)=
-\tfrac{1}{N}\sum\nolimits_i(\sigma_iw_i\phi - \ln\cosh w_i\phi),
\end{equation}
such that 
\begin{equation}
P(\bm\sigma\mid\phi)=\prod\nolimits_iP(\sigma_i\mid\phi)=e^{-N g(\phi;\bm\sigma)}/2^N.
\end{equation}
Therefore the full joint distribution can be written as
\begin{equation}\label{eq:fulljointdist}
P(\bm\sigma,\phi)
=
P(\bm\sigma\mid\phi)P(\phi)
=
P(\phi)\times e^{-Ng(\phi;\bm\sigma)}/2^N.
\end{equation}

We now consider the limit $N\!\to\!\infty$. We assume that the weights $\{w_i\}$ are independent of the system size $N$ (e.g., they are drawn from a fixed distribution) such that $g(\phi;\bm\sigma)$ is intensive. For a \emph{smooth} prior---i.e., $\lim_{N\to\infty}\frac{1}{N}\ln P(\phi)\!=\!0$---the joint distribution [Eq.~\eqref{eq:fulljointdist}] is dominated by fluctuations around the minimum of $g(\phi;\bm\sigma)$, 
\begin{equation}
P(\bm\sigma,\phi)
\approx
P(\phi^*_{\bm\sigma})
P(\bm\sigma\mid\phi^*_{\bm\sigma})\times
e^{-\frac{N}{2}g''(\phi^*_{\bm\sigma})(\phi-\phi^*_{\bm\sigma})^2}
\!\!,
\end{equation}
where $\phi^*_{\bm\sigma}$ is the root of $g'(\phi;\bm\sigma)\!=\!0$ and we drop the superfluous dependence on $\bm\sigma$ from $g''(\phi;\bm\sigma)\!=\!\tfrac{1}{N}\sum\nolimits_iw_i^2\sech^2 w_i\phi$. We see that when conditioned on the spins, the fluctuation is Gaussian,
\begin{equation}\label{eq:P_phi_sigma}
P(\phi\mid\bm\sigma)
\approx
\mathcal{N}\left(
\phi
\mid
\mu\!=\!\phi^*_{\bm\sigma}
,\ 
s^2\!=\!\frac{1}{Ng''(\phi^*_{\bm\sigma})}
\right).
\end{equation}
As a result, we obtain the conditional differential entropy,
\begin{equation}\label{eq:S_phi_sigma}
S(\phi\mid\bm\sigma)
\approx
-\frac{1}{2}\ln N
+\frac{1}{2}\ln 2\pi e
-\frac{1}{2}\sum_{\bm\sigma} P(\bm\sigma)\ln g''(\phi^*_{\bm\sigma}).
\end{equation}
We see that the logarithmic divergence is the leading contribution since the last two terms do not grow with $N$. For extrinsic fluctuations, $P(\phi)$, and thus $S(\phi)$, is independent of $N$; therefore, we obtain 
\begin{equation}
I(\phi;\bm\sigma)=
S(\phi)-S(\phi\mid\bm\sigma)
\approx
\frac{1}{2}\ln N
+O(N^0).
\end{equation}
On the other hand, if $P(\phi)\!\sim\!e^{-N^\gamma \times c(\phi-\phi_0)^2}$ for some $\gamma\!\in\![0,1)$ and a constant $c\!>\!0$, its entropy diverges logarithmically $S(\phi)\!\approx\!-\frac{\gamma}{2}\ln N+O(N^0)$ and the mutual information reads
\begin{equation}
I(\phi;\bm\sigma)
\approx
\frac{1-\gamma}{2}\ln N
+O(N^0).
\end{equation}
We see that the decrease in information results from the fluctuation entropy that decreases logarithmically with $N$.

The same logarithmic divergence also emerges in the mutual information between two macroscopic halves of the system. To see this, we note that the entropy of the spins is given by
\begin{equation}
S(\bm\sigma)
=S(\phi)-S(\phi\mid\bm\sigma) + S(\bm\sigma\mid\phi)
\end{equation}
Similarly, the entropy of each half of the system reads
\begin{equation}
S(\bm\sigma^\nu)
=S(\phi)-S(\phi\mid\bm\sigma^\nu) + S(\bm\sigma^\nu\mid\phi)
\end{equation}
where $\nu\!\in\!\{A,B\}$ and $\bm\sigma\!=\!(\bm\sigma^A,\bm\sigma^B)$. 
When the subsystems are large and the spins in each half are randomly chosen, we have [see Eq.~\eqref{eq:S_phi_sigma}]
\begin{equation}
S(\phi\mid\bm\sigma^\nu)
\approx
S(\phi\mid\bm\sigma)
-\frac{1}{2}\ln\frac{N_\nu}{N}.
\end{equation}
We now write the mutual information between the two halves in terms of the above entropy,
\begin{align}
I(\bm\sigma^A;\bm\sigma^B) 
&= S(\bm\sigma^A)+S(\bm\sigma^B)-S(\bm\sigma)
\\
&\approx I(\phi;\bm\sigma) +\frac{1}{2} \ln \frac{N_AN_B}{N^2}.
\end{align}
where we use the fact that $I(\phi;\bm\sigma)\!=\!S(\phi)\!-\!S(\phi\,|\,\bm\sigma)$ and the property of conditional independence, $S(\bm\sigma\,|\,\phi)=S(\bm\sigma^A\,|\,\phi)+S(\bm\sigma^B\,|\,\phi)$. For $N_A\!=\!N_B\!=\!N/2$, we see that $I(\bm\sigma^A;\bm\sigma^B)$ is smaller than $I(\phi;\bm\sigma)$ by one bit.

\section{Rank-one Ising models\label{appx:rank1}}

Here we provide an analysis of rank-one Ising models---those with pairwise interaction matrices of rank one---defined by the energy function,
\begin{equation}
    \H(\bm\sigma)
    =
    -\frac{1}{4N}\sum\nolimits_{ij}w_iw_j\sigma_i\sigma_j
    =
    -\frac{1}{4N}\left(\sum\nolimits_{i}w_i\sigma_i\right)^2,
\end{equation}
where $\bm\sigma\!=\!(\sigma_1,\sigma_2,\dots,\sigma_N)$ denotes the state of the system, $\sigma_i\!\in\!\{\pm1\}$ the spin at site $i\!\in\!\{1,2,\dots,N\}$, and the product $w_iw_j$ describes the interaction between spins $i$ and $j$. We note that the terms with $i\!=\!j$ only add an irrelevant constant. This model generalizes the fully-connected Ising model which corresponds to setting $w_i\!=\!w_j$ for all $i$ and $j$. As usual, the probability distribution of the system configuration is given by
\begin{equation}
    P(\bm\sigma) = e^{-\beta \H({\bm\sigma})}/Z,
\end{equation}
where we introduce the inverse temperature $\beta\!=\!1/T$ and the partition function $Z\!=\!\sum_{\bm\sigma}e^{-\beta \H(\bm\sigma)}$.

Computing the partition function by directly summing over all possible spin states is generally analytically intractable. Instead, we trade this summation for an integral using the Hubbard--Stratonovich transformation,
\begin{equation}\label{eq:hubbardstratonovich}
    Z 
    = \sum_{\bm\sigma}e^{\frac{\beta}{4N}\left(\sum_{i}w_i\sigma_i\right)^2}
    = 
    \sqrt{\frac{N}{\pi\beta}}\int\!\!\! d\phi 
    \sum_{\bm\sigma}
    e^{-\frac{N}{\beta}\phi^2 +\phi \sum_iw_i\sigma_i}.
\end{equation}
We see that the spins become noninteracting at the cost of introducing a new fluctuating field $\phi$ which correlates with the spins via the joint distribution,
\begin{equation}\label{eq:jointdist}
    P(\bm\sigma,\phi)
    =
    \frac{1}{Z}
    \sqrt{\frac{N}{\pi\beta}}
    e^{-\frac{N}{\beta}\phi^2 +\phi \sum_iw_i\sigma_i}.
\end{equation}
Summing out each spin variable from Eq.~\eqref{eq:hubbardstratonovich} yields
\begin{equation}\label{eq:Z}
    Z
    =
    2^N
    \sqrt{\frac{N}{\pi\beta}}
    \int\!\!\! d\phi \,
    e^{-\frac{N}{\beta}\phi^2 + \sum_i \ln\cosh w_i\phi}.
\end{equation}
As a result, we can express various thermodynamic variables of the spins as integrals over a continuous field which are usually more convenient than summations over discrete spin states. In particular, the internal energy, entropy and heat capacity---$U$, $S$ and $C$, respectively---read
\begin{align}
    \label{eq:U}
    U&
    = 
    -\frac{\partial}{\partial\beta}\ln Z
    =
    \frac{1}{2\beta}
    \left(
    1
    -
    \frac{2N}{\beta}\langle\phi^2\rangle
    \right)
    \\
    \label{eq:S}
    S&
    = 
    \beta U + \ln Z
    \\
    C&
    =
    \beta^2\frac{\partial^2}{\partial\beta^2}\ln Z
    =
    2\beta U
    -
    \frac{1}{2}
    +
    \frac{N^2}{\beta^2}
    \var(\phi^2),
\end{align}
where $\langle\phi^2\rangle$ and $\var(\phi^2)$ denote the mean and variance of $\phi^2$.

In addition, we see that a rank-one Ising model is equivalent to a conditionally independent model with a fluctuation field, induced by the intrinsic interactions between spins. The marginal distribution of this field is given by
\begin{equation}\label{eq:latentprior}
    P_N(\phi)
    =
    \sum_{\bm\sigma}P(\bm\sigma,\phi)
    =
    \frac{2^N}{Z}
    \sqrt{\frac{N}{\pi\beta}}
    e^{-\frac{N}{\beta}\phi^2 + \sum_i \ln\cosh w_i\phi}, 
\end{equation}
and thus we have
\begin{equation}
    P(\bm\sigma\mid\phi)
    =
    \frac{P(\bm\sigma,\phi)}{P_N(\phi)}
    =
    \prod_i
    \frac{e^{\sigma_iw_i\phi}}{2\cosh w_i\phi}.
\end{equation}
We see again that conditioning on the fluctuating field removes the interactions between spins.

Recalling that $\ln\cosh x \approx\frac{x^2}{2}\!-\!\frac{x^4}{12}$ for small $x$, we see that the fluctuation distribution $P_N(\phi)$ exhibits a structural transition at 
\begin{equation}
    \beta_c = \frac{2}{\frac{1}{N}\sum_i w_i^2} 
\end{equation}
where it changes from unimodal at $\beta\!<\!\beta_c$ to bimodal at $\beta\!>\!\beta_c$. In the limit $N\!\to\!\infty$, this point corresponds to the critical temperature of an order-disorder phase transition. In the disordered phase at high temperatures $\beta\!<\!\beta_c$, the spins are mostly random. In the ordered phase at low temperatures $\beta\!>\!\beta_c$, on the other hand, they mimic the pattern set by the signs of $\{w_i\}$.

%
\end{document}